\documentclass[12pt]{article}
\usepackage{graphicx}

\newcommand\pubnumber{}
\newcommand\pubdate{\today}

\def\napoli{Department of Physics\\
University of Oxford, Keble Road, Oxford OX13RH, United Kingdom}

\def\Title#1{\begin{center} {\Large #1 } \end{center}}
\def\Author#1{\begin{center}{ \sc #1} \end{center}}
\def\Address#1{\begin{center}{ \it #1} \end{center}}

\newcommand\pubblock{\rightline{\begin{tabular}{l} \pubnumber\\
         \pubdate  \end{tabular}}}
\newenvironment{Abstract}{\begin{quotation}  }{\end{quotation}}
\newenvironment{Presented}{\begin{quotation} \begin{center} 
             PRESENTED AT\end{center}\bigskip 
      \begin{center}\begin{large}}{\end{large}\end{center} \end{quotation}}

\def\beq{\begin{equation}}
\def\eeq#1{\label{#1}\end{equation}}
\def\eeqn{\end{equation}}

\def\beqa{\begin{eqnarray}}
\def\eeqa#1{\label{#1}\end{eqnarray}}
\def\eeqan{\end{eqnarray}}

\let\bar=\overbar

\def\Dslash{\not{\hbox{\kern-4pt $D$}}}
\def\dslash{\not{\hbox{\kern-2pt $\del$}}}

\def\msb{{\bar{\ssstyle M \kern -1pt S}}}

\def\Journal#1#2#3#4{{#1} {\bf #2}, #3 (#4)}

\begin{document}
\begin{titlepage}
\pubblock

\vfill
\Title{Lifetime and mixing parameters of neutral D mesons and neutral B hadrons in experimental particle physics.}
\vfill
\Author{ Sneha Malde}
\Address{\napoli}
\vfill
\begin{Abstract}
A review of recent experimental measurements of the lifetime and mixing parameters of neutral D and B hadrons is presented. In particular, focus is given to measurements of D mixing and the $\Lambda_B$ lifetime. 
\end{Abstract}
\vfill
\begin{Presented}
Proceedings of CKM 2010, the 6th international workshop on the CKM unitary triangle, University of Warwick, Sept 6-10th 2010.
\end{Presented}
\vfill
\end{titlepage}
\def\thefootnote{\fnsymbol{footnote}}
\setcounter{footnote}{0}

\section{D Meson Mixing}

The weak eigenstates of neutral mesons are different to their mass eigenstates. This leads to the phenomenon of mixing whereby neutral mesons oscillate between their matter and antimatter state. This was observed first in the Kaon sector and subsequently in $B^0$ and $B_s$ mesons. We define the mixing parameters $x$ and $y$ in terms of the masses and the widths of the two eigenstates as $x = (M_1-M_2)/\Gamma$ and $y=(\Gamma_1 - \Gamma_2)/2\Gamma$. In $D^0$ mesons $x$ and $y$ are expected to be very small~\cite{smallMix} and a number of analyses have tried to measure them. CP violation is also expected to be small in D meson mixing~\cite{smallCPV}. The majority of analysis can be extended to look for CPV, but these are not covered here. A full review is given here~\cite{PDG}. Examples of two interesting measurements are given below.

\subsection{Wrong Sign decay analysis}

One of the first measurements to yield evidence of $D^0$ mixing is the ``wrong sign'' analysis. To determine mixing the production flavor must be known. This is achieved by using $D^0$ mesons that are the daughters of $D^*$ decays. The decay $D^{*+}\to D^0\pi^+$ tags the initial flavor of the $D$ by the charge of the slow pion. The dominant or ``right sign'' decay is $D^0 \to K^-\pi^+$. The wrong sign decay to $K^+,\pi^-$ occurs via two possible processes. The decay can occur via a doubly Cabibbo suppressed decay, or it can occur via mixing followed by the Cabibbo favoured decay of the $\bar{D^0}$. There is interference between the two processes. The number of wrong sign decays as a function of time is given by
\begin{equation}
\frac{dN_{WS}}{dt}(t) \propto e^{-\Gamma t}\cdot[R_d + y'(\Gamma t)\sqrt{R_D} + \frac{x'^2 +y'^2}{4}(\Gamma t)^2]
\end{equation}
where the three terms in the square parentheses represent the contributions of the doubly Cabibbo suppressed, interference and mixing contributions respectively, and $R_D$ is the ratio of suppressed to favoured decay amplitudes. The expression is sensitive to $x'$ and $y'$ which are related to $x$ and $y$ via the strong phase $\delta^{K\pi}$ such that $x' = x \mathrm{cos}(\delta^{K\pi}) + y \mathrm{sin}(\delta^{K\pi})$ and $ y' = x\mathrm{cos}(\delta^{K\pi}) - y \mathrm{sin}(\delta^{K\pi})$.

\begin{figure}
\centering
\includegraphics[height=2.2in]{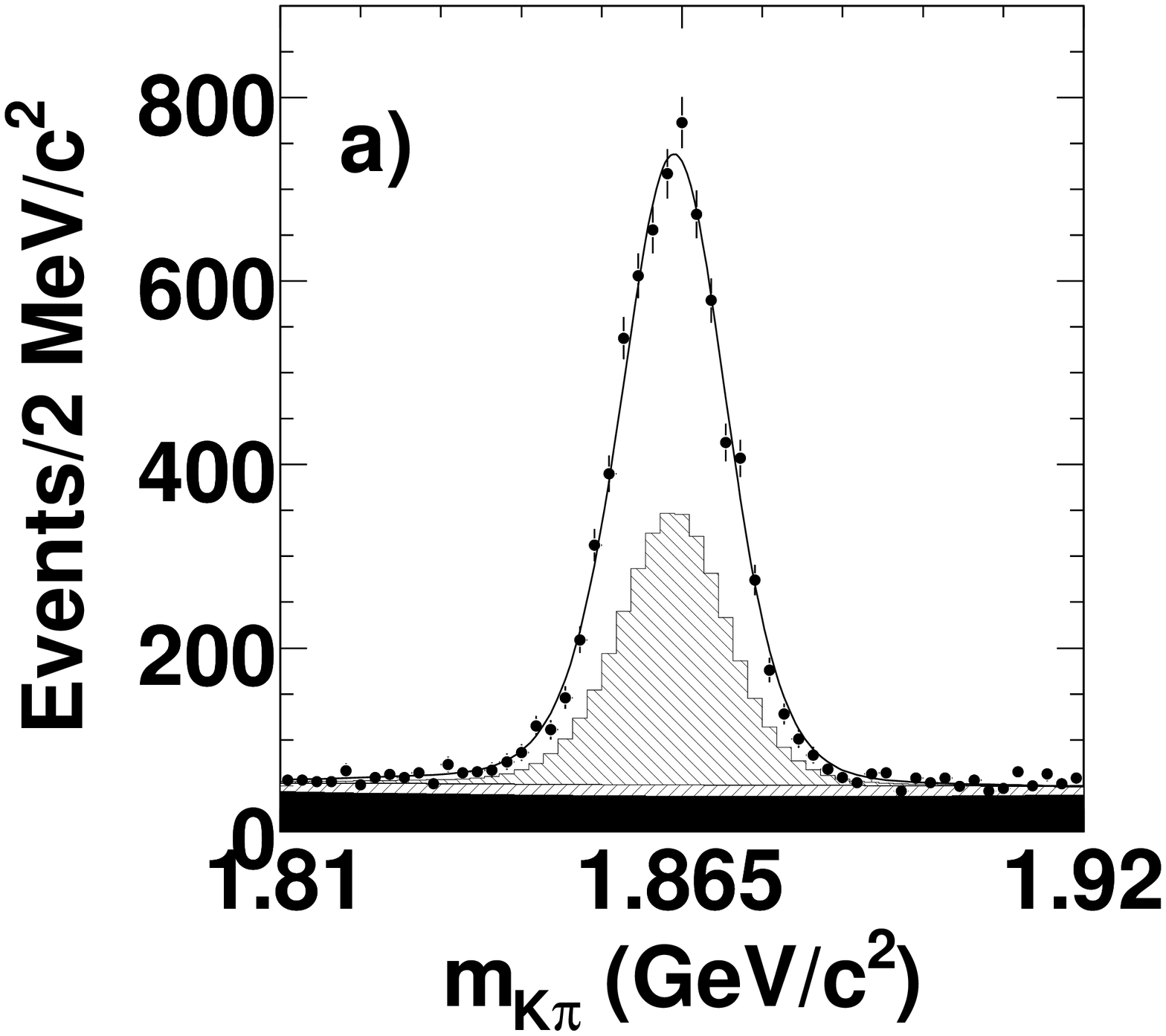}
\includegraphics[height=2.2in]{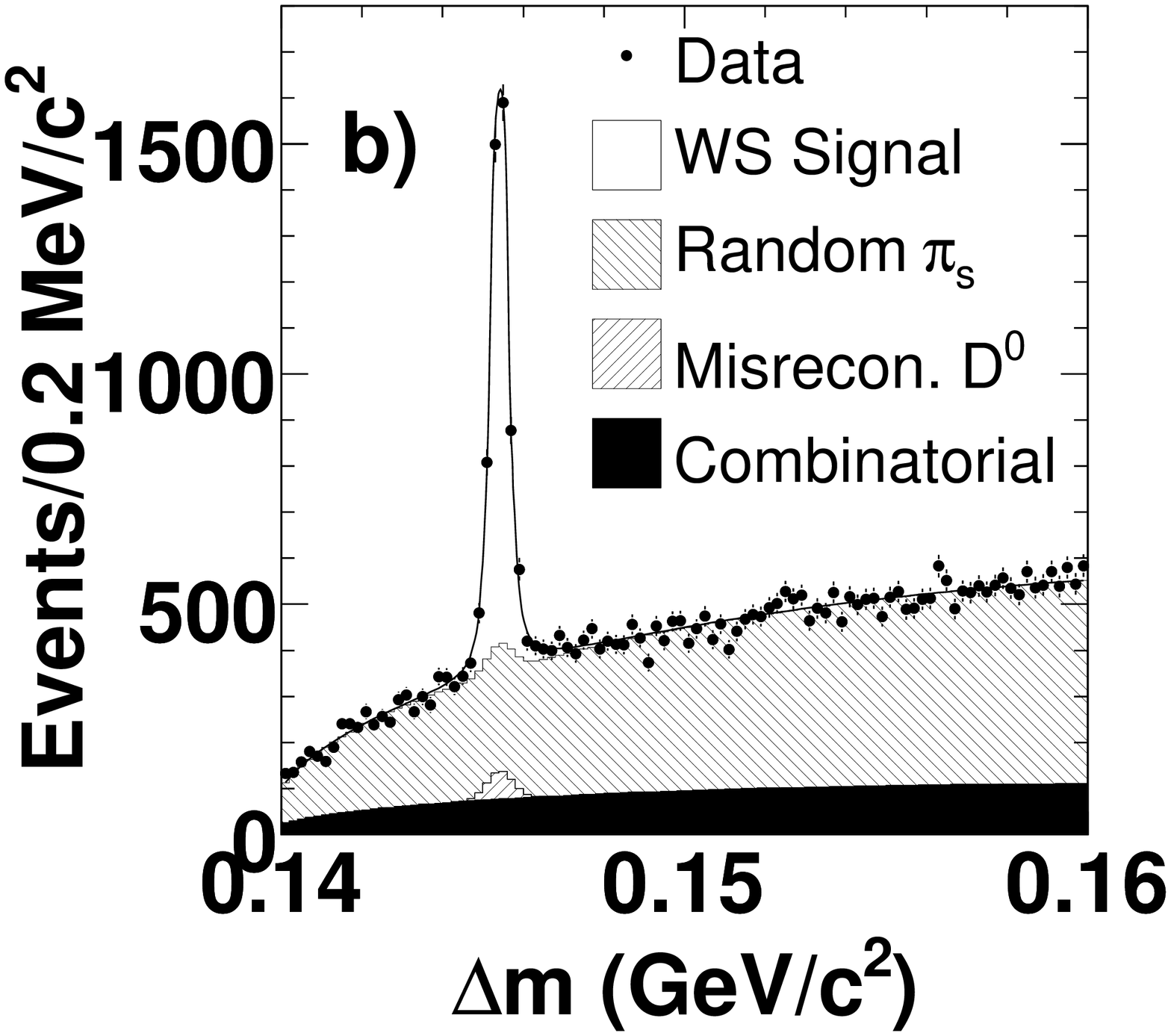}
\caption{
\label{dmassfit} These plots are from the Babar analysis ~\cite{BabWS}. The left plot shows the fit to the $D^0$ mass and the right to the mass difference between the $D^*$ and the $D^0$. The contribution of different backgrounds is shown.}
\end{figure}

An analysis of wrong sign decays has been carried out at Babar~\cite{BabWS}, Belle~\cite{BelleWS} and subsequently at CDF~\cite{CDFWS}. Important detector requirements are accurate measurements of productions and decay vertices, a way to identify the tracks as either Kaons or Pions, and also to account for background where the $D^*$ is itself a secondary particle. Taking the Babar analysis as n example, the sample composition is determined from fits to the mass of the $D^0$ and also the mass difference between the reconstructed $D^0$ and $D^*$. Both distributions are analysed as some backgrounds are flat in one distribution but peaking in the other, as shown in Fig.~\ref{dmassfit}.

\begin{figure}
\centering
\includegraphics[height=2.2in]{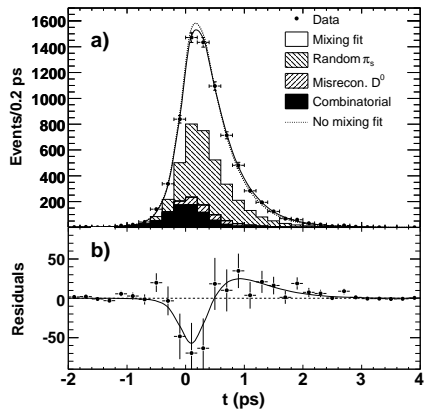}
\caption{
\label{mixing}  This plot is from the Babar analysis ~\cite{BabWS}. Proper time distribution of data. The residual between the mixing/no mixing fit and data is shown.}
\end{figure}

The decay time distribution of the right sign sample is fitted to a single exponential convolved by a decay time resolution function. The determined resolution is then fixed in the wrong sign decay time fit. The fit is performed with and without the mixing terms included in the likelihood. From Fig.~\ref{mixing} it is seen that the data prefer the mixing hypothesis. The result of the Babar analysis is that the best fit point for $x^{2'}$ and $y'$ is displaced from (0,0), the no mixing point, by 3.9$\sigma$. The analysis of the same channel at CDF and Belle disfavour the no-mixing hypothesis at 3.8$\sigma$ and 2.0$\sigma$ respectively.


\subsection{Direct measurements of $x$ and $y$}

A recent measurement of D mixing allows for a direct measurement of $x$ and $y$. The analysis involves a time dependent amplitude analysis of $D^0 \to K_s hh$, where $h = \pi, K$. The analysis involves studying the distribution of events across Dalitz space as a function of decay time. An observed variation in the Dalitz space distribution as a function of time is a signature of mixing. The sensitivity to $x$ and $y$ comes mainly from regions with interference of Cabibbo favoured and doubly Cabibbo suppressed decays or from regions where the decay passes through a CP eigenstate. A simultaneous amplitude and decay time fit is performed to determine $x$ and $y$. The results are $x=(0.16 \pm 0.23 \pm 0.12 \pm 0.08)\%)$ and $y=(0.57\pm 0.20\pm 0.13\pm 0.07)\%)$ where the first uncertainty is statistical, the second systematic and the third relates to the amplitude model used. This is the most precise single measurement of $x$ and the results disfavour the no mixing hypothesis at 1.9 sigma~\cite{KspipiBab}. The results are consistent with a similar analysis performed at Belle~\cite{KspipiBelle}.

Although no single results disfavours the no-mixing hypothesis at more than 5$\sigma$ the combination of all results disfavours no-mixing at more than 10$\sigma$.

\section{B hadron lifetimes}

The experimental measurement of $B$ hadron lifetime ratios is an important test
of the theoretical approach to $B$ hadron observables known as the Heavy Quark Expansion~\cite{HQE} (HQE).  The lifetime of ground-state hadrons containing a $b$ quark and lighter quarks is largely determined by the charged weak decay of the $b$ quark. Corrections include interactions of the light quarks, Pauli interference, weak annihilation and weak scattering and are collected together in the heavy quark expansion in orders of $\Lambda_{QCD}/m_b$. These other effects alter the lifetimes at approximately the 10\% level. While precise predictions for $B$ hadron lifetimes are difficult to calculate the ratios are predicted with fairly high accuarcy by the HQE. This framework of theoretical calculation is used to predict low energy QCD effects in many flavour observables. For example, HQE predicts the decay width of $B_s$ mesons to final states common to $B^0_s$ and $\bar{B^0_s}$ ~\cite{HQEDG}, $\Gamma_{12}^s$, which enters the decay-width difference in the $B^0_s$ system and several $CP$ violation effects. The measurement of lifetime ratios provides a simple and accurate way to test the HQE framework as non standard model effects are expected to be highly suppressed in lifetimes.

 The ratio $\tau(B^+)/\tau(B^0)$ (charge conjugates are implied throughout) is predicted~\cite{NLOQCD,Bigi,Subleading} to be in the range 1.04--1.08 and the ratio $\tau(\Lambda^0_b)/\tau(B^0)$ in the range 0.83--0.95~\cite{NLOQCD, Subleading, LBOriginals}. The measured world average $B^+$ and $B^0$ lifetimes are dominated by the Belle experiment~\cite{ref:BelleResult}. Of recent interest is the $\Lambda^0_b$ lifetime. Until 2006 all measurements were in agreement but lay at the lower end of the theoretically expected value. Since then, two high precision CDF measurements are significantly above previous results~\cite{ref:Mark, ref:Petar}. The analysis described here is the most precise measurement of the $B^+$, $B^0$, and $\Lambda^0_b$ lifetimes and ratios. 

The $B^+$, $B^0$, and $\Lambda_b$ lifetimes have been measured using the decay channels $B^+ \to J/\psi K^+$, $B^0 \to J/\psi K^*$, $B^0 \to J/\psi K_{s}$ and $\Lambda_b \to J/\psi \Lambda$ in $4.3fb^{-1}$ of data collected by the CDF detector~\cite{lifetimes}. To control systematics uncertainties the strategy employed is to use similar selection criteria. The selection uses rectangular cuts on time-independent variables chosen to maximise $S/\sqrt{(S+B)}$, where $S$ and $B$ are the numbers of signal and background events respectively. The observed yields are $B^+:$ 45000$\pm$230, $B^0\to J/\psi K^*:$16860 $\pm$140,  $B^0\to J/\psi K_s:$12070 $\pm$120, and $\Lambda_b :$ 1710 $\pm$ 50. The majority of the background events are combinatorial.

\begin{figure}
\centering
\includegraphics[height=1.8in]{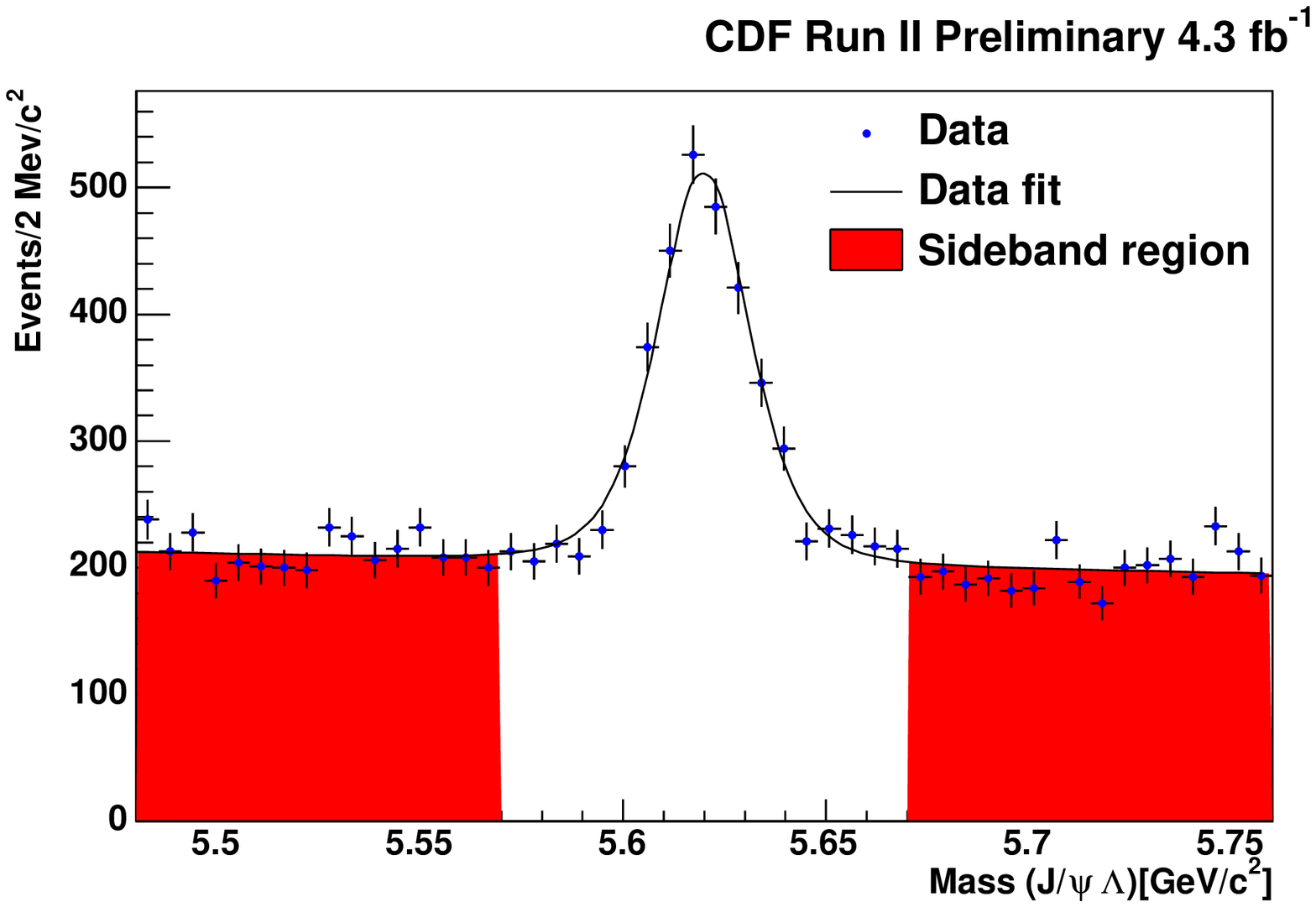}
\includegraphics[height=1.8in]{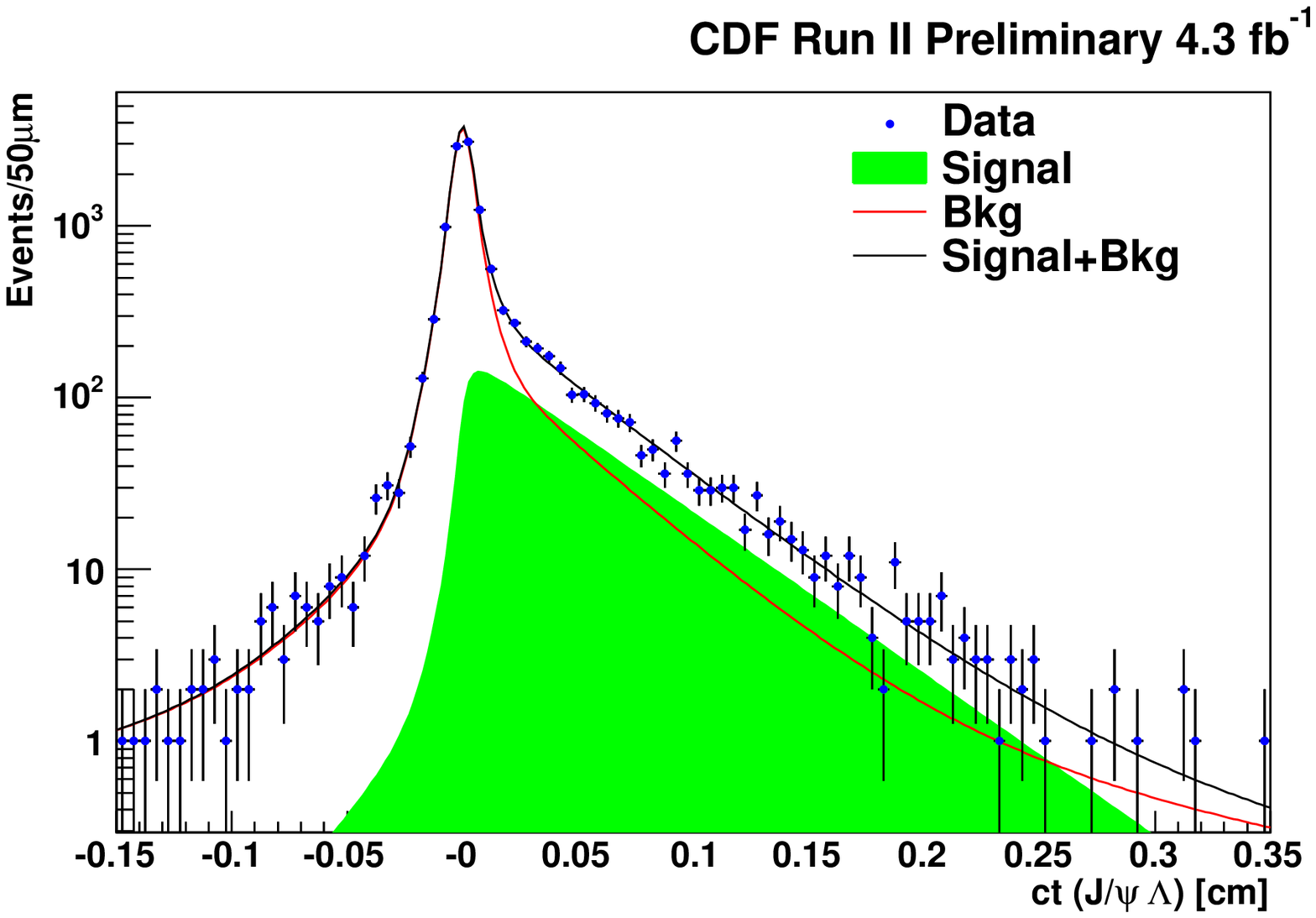}
\caption{These plots are from ~\cite{lifetimes}. The $\Lambda_b$ mass and proper decay time distributions. Fit projections are also shown.
\label{blifes}}
\end{figure}

In previous measurements from CDF the uncertainty due to detector resolution has been a leading source of systematic uncertainty. The proper decay time is determined using the $J/\psi$ vertex to provide similarity in the decay time resolution between channels and to allow for the cancellation of certain systematic uncertainties. A detailed resolution model is also introduced in this analysis. The parameters of the resolution function are determined from the mass sidebands as the fraction of background events expected to originate from the primary vertex is between 80-90$\%$, depending on channel and background model, and therefore provides a useful sample from which to determine the resolution. The overall fit is an unbinned likelihood fit to the mass, decay time and decay time uncertainty distributions simultaneously. The projections of the mass and decay time distributions of the $\Lambda_b$ are shown in Fig.~\ref{blifes}. 

The best fit lifetimes are $\tau_{B^+} = 1.639 \pm 0.009 ~({\rm stat}) \pm 0.009~{\rm (syst)~ ps}$, $\tau_{B^0} = 1.507 \pm 0.010 ~({\rm stat}) \pm 0.008~{\rm (syst)~ ps}$, and $\tau_{\Lambda^0_b} = 1.537 \pm 0.045 ~({\rm stat}) \pm 0.014~{\rm (syst)~ ps}$. The lifetime ratios are calculated as  $\tau_{B^+}/\tau_{B^0} = 1.088  \pm 0.009~({\rm stat})\pm 0.004~({\rm syst})$ and $\tau_{\Lambda^0_b}/\tau_{B^0}= 1.020   \pm 0.030~({\rm stat})\pm 0.008~({\rm syst})$~\cite{lifetimes}. These are the world's best measurements of the lifetimes and ratios. The $B^+$ and $B^0$ lifetime and their ratio are consistent with previous measurements. In particular the ratio has high precision as the leading systematic uncertainties are correlated between channels and hence cancel in the ratios. The precision is more accurate than the theoretical uncertainties and could be used in the future for comparison to lattice QCD calculations. The improvement in the systematic uncertainty from 0.033 ps (1.0 fb$^{-1}$) to 0.014 ps (4.3 fb$^{-1}$) is evident in the $\tau(\Lambda_b)$ measurement. The $\Lambda_b$ lifetime remains higher than the world average but is not inconsistent with theoretical predictions. Further measurements, in particular from the LHCb experiment are awaited to resolve the true $\Lambda_b$ lifetime. More accurate measurements and predictions will be required to understand if there is any tension between the measured $\tau_{\Lambda^0_b}/\tau_{B^0}$ ratio and the HQE prediction.  

The $B_s$ meson has two mass eigenstates, leading to two widths, $\Gamma_H$ and $\Gamma_L$. One further crucial test of HQE is the lifetime ratio of the $B_s$ and $B_d$ mesons, as due to their similarity the uncertainty on the theoretical prediction, $\tau_{B_s}/\tau_{B^0} = 1.0  \pm 0.01 $~\cite{NLOQCD}, is small. The average lifetime of these two eigenstates, $\frac{2}{\Gamma_H + \Gamma_L}$ is best measured through the combined time and angular analysis of $B_s \to J/\psi \phi$. This analysis, measures $\bar{\tau_{B_s}}= 1.53 \pm 0.025 (\rm{stat}) \pm 0.012 (\rm{syst}) ps$ ~\cite{bsCDF} at CDF and  $\bar{\tau_{B_s}}= 1.45 \pm 0.04 (\rm{stat}) \pm 0.01 (\rm{syst}) ps$~\cite{bsD0} at D0. These measurements are consistent with the theoretical prediction. The width difference is measured to be $\Delta \Gamma_s = 0.075 \pm 0.035(\rm{stat}) \pm 0.01(\rm{syst}) ps^{-1}$ and $\Delta \Gamma_s = 0.15 \pm 0.06(\rm{stat}) \pm 0.01(\rm{syst}) ps^{-1}$ at CDF and D0 respectively. These are also consistent with the theoretical prediction of $\Delta \Gamma_s = 0.070 \pm 0.042 \rm{ps^{-1}}$~\cite{HQEDG}.

\section{Summary}

Recent lifetime measurements remain consistent with HQE and give confidence in this theoretical framework. The new $\Lambda_b$ lifetime measurement remains higher than previous measurements and anticipated new measurements from LHCb will be required to resolve the tension here. Since the first evidence of $D^0$ mixing in 2007 a number of analyses have been used to search for evidence of mixing. While no single measurement has observed it, the sum of the measurements overwhelmingly indicates that the values of $x$ and $y$ are not consistent with 0. Future measurements from LHCb in particular are waited to improve knowledge on D mixing, lifetime differences in the $D$ and $B$ mesons and the $\Lambda_b$ lifetime.



\end{document}